\documentclass[fleqn,usenatbib]{mnras}

\usepackage{newtxtext,newtxmath}

\usepackage[T1]{fontenc}

\DeclareRobustCommand{\VAN}[3]{#2}
\let\VANthebibliography\thebibliography
\def\thebibliography{\DeclareRobustCommand{\VAN}[3]{##3}\VANthebibliography}

\usepackage{graphicx}	
\usepackage{amsmath}	

\newcommand{\Ni}{\ensuremath{^{56}\mathrm{Ni}}}

\newcommand{\Msun}{\ensuremath{\mathrm{M}_\odot}}

\graphicspath{{./}{figures/}}

\title[LC bumps from variable magnetar thermal energy injection]{
Variable thermal energy injection from magnetar spin down as a possible cause of stripped-envelope supernova light-curve bumps
}

\author[T. J. Moriya et al.]{
Takashi J. Moriya,$^{1,2}$\thanks{E-mail: takashi.moriya@nao.ac.jp (TJM)}
Kohta Murase,$^{3,4,5,6}$
Kazumi Kashiyama,$^{7,8}$
and 
Sergei I. Blinnikov$^{9,10}$
\\
$^{1}$National Astronomical Observatory of Japan, National Institutes of Natural Sciences, 2-21-1 Osawa, Mitaka, Tokyo 181-8588, Japan \\
$^{2}$School of Physics and Astronomy, Faculty of Science, Monash University, Clayton, Victoria 3800, Australia \\
$^{3}$Department of Physics, The Pennsylvania State University, University Park, PA 16802, USA \\
$^{4}$Department of Astronomy and Astrophysics, The Pennsylvania State University, University Park, PA 16802, USA \\
$^{5}$Center for Multimessenger Astrophysics, Institute for Gravitation and the Cosmos, The Pennsylvania State University, University Park, PA 16802, USA \\
$^{6}$Center for Gravitational Physics, Yukawa Institute for Theoretical Physics, Kyoto University, Sakyo, Kyoto 606-8502, Japan \\
$^{7}$Research Center for the Early Universe, Graduate School of Science, University of Tokyo, Bunkyo, Tokyo 113-0033, Japan \\
$^{8}$Department of Physics, Graduate School of Science, University of Tokyo, Bunkyo, Tokyo 113-0033, Japan \\
$^{9}$NRC Kurchatov Institute, 123182 Moscow, Russia \\
$^{10}$Dukhov Automatics Research Institute (VNIIA),  127055 Moscow, Russia
}

\date{Accepted 2022 May 11. Received 2022 April 19; in original form 2022 February 7}

\pubyear{2022}

\begin{document}
\label{firstpage}
\pagerange{\pageref{firstpage}--\pageref{lastpage}}
\maketitle

\begin{abstract}
Luminosity evolution of some stripped-envelope supernovae such as Type~I superluminous supernovae is difficult to be explained by the canonical \Ni\ nuclear decay heating. A popular alternative heating source is rapid spin down of strongly-magnetized rapidly-rotating neutron stars (magnetars).
Recent observations have indicated that Type~I superluminous supernovae often have bumpy light curves with multiple luminosity peaks. The cause of bumpy light curves is unknown. In this study, we investigate the possibility that the light-curve bumps are caused by variations of the thermal energy injection from magnetar spin down. We find that a temporal increase in the thermal energy injection can lead to multiple luminosity peaks. The multiple luminosity peaks caused by the variable thermal energy injection is found to be accompanied by significant increase in photospheric temperature, and photospheric radii are not significantly changed. We show that the bumpy light curves of SN~2015bn and SN~2019stc can be reproduced by temporarily increasing magnetar spin-down energy input by a factor of $2-3$ for $5-20$~days. However, not all the light-curve bumps are accompanied by the clear photospheric temperature increase as predicted by our synthetic models. In particular, the secondary light-curve bump of SN~2019stc is accompanied by a temporal increase in photospheric radii rather than temperature, which is not seen in our synthetic models. We, therefore, conclude that not all the light-curve bumps observed in luminous supernovae are caused by the variable thermal energy injection from magnetar spin down and some bumps are likely caused by a different mechanism. 
\end{abstract}




\section{Introduction}
Supernovae (SNe) are explosions of stars that can be as luminous as galaxies. There are several heating mechanisms that make SNe luminous. The major heating source of stripped-envelope SNe, which are explosions of stripped stars with little or no hydrogen-rich envelopes, is considered to be the nuclear decay energy of the radioactive \Ni\ synthesized during their explosions \citep[][]{pankey1962,arnett1982}. However, recent transient surveys are starting to discover SNe that are difficult to be powered by the \Ni\ decay heating \citep[][]{kasen2017}. 
For example, the luminosity decline of fast blue optical traisients (FBOTs, see \citealt{inserra2019} for a recent review) is much faster than that expected from the \Ni\ decay heating \citep[][and references therein]{ho2021}. Another example is Type~I superluminous SNe (SLSNe, \citealt{quimby2011}) which sometimes require more than 10~\Msun\ of \Ni\ to explain their peak luminosity \citep[e.g.,][]{gal-yam2009,nicholl2016sn2015bn1,umeda2008}. SLSNe usually do not show observational signatures expected when a large amount of \Ni\ is produced \citep[e.g.,][]{dessart2012,jerkstrand2016,mazzali2019}. Thus, their luminosity source is not considered to be the \Ni\ decay heating \citep[e.g.,][for a review]{moriya2018slsn}. It has also been suggested that some luminous stripped-envelope SNe may not be powered by the \Ni\ heating even if they are not as luminous as SLSNe \citep[e.g.,][]{kashiyama2016,afsariardchi2021,sollerman2021}.

There are several alternative heat sources to account for the stripped-envelope SNe whose luminosity evolution cannot be explained by the canonical \Ni\ heating. One is the interaction between SN ejecta and dense circumstellar matter (CSM, e.g., \citealt{sorokina2016}). When SN ejecta hit a dense CSM, the kinetic energy of the SN ejecta can be efficiently converted to thermal energy that can provide additional heat to illuminate SNe. Indeed, some stripped-envelope SNe such as Type~Ibn and Type~Icn SNe show observational signatures of dense CSM surrounding the SN ejecta \citep[e.g.,][]{pastorello2007,ben-ami2014,moriya2016,hosseinzadeh2017,gangopadhyay2020,fraser2021,perley2021,gal-yam2021,maeda2022}.

Another possible alternative heat source is a central compact remnant formed during a SN explosion. A popular mechanism of this kind is rapid spin down of strongly-magnetized rapidly-rotating neutron stars \citep[][]{ostriker1971,gaffet1977,maeda2007,kasen2010,woosley2010}. Such a neutron star is often called a ``magnetar'' in luminous SN studies. The huge rotational energy of magnetars can be released in a spin-down timescale of less than 100~days if their magnetic fields are more than $\sim 10^{13}~\mathrm{G}$, making SNe as luminous as SLSNe \citep[e.g.,][]{barkov2011,inserra2013,metzger2015,nicholl2017}. Even if a black hole is formed at the center, alternatively, accretion towards the central black hole can be a heating source to make SNe luminous \citep[e.g.,][]{dexter2013,moriya2018}.

Most stripped-envelope SNe have been characterized by a single peaked light curves (LCs) and their luminosity evolution is usually reproduced by assuming a single heating source. However, it is starting to be realized that the LCs of some stripped-envelope SNe have bumpy structure with multiple luminosity peaks \citep[e.g.,][]{maeda2007,roy2016,yan2017,yan2020,gomez2021,prentice2021}. Especially, SLSNe are suggested to have bumpy LCs generally \citep[][]{hosseinzadeh2021,chen2022a,chen2022b}. LC bumps in SLSNe are found to be diverse. For example, some SLSNe have a smooth round-peak bump \citep[e.g., SN~2019stc;][]{gomez2021}, while some have a triangle-shaped bump \citep[e.g., PS1-12cil;][]{lunnan2018}. Even if the LC bump does not clearly exist, some extra luminosity contribution is suggested to be required to explain spectral properties of some SLSNe \citep[e.g.,][]{chen2017}. The origin of the multiple luminosity peaks is still unknown. It is often assumed that each bump is caused by a different heating source \citep[e.g.,][]{li2020}. For instance, a combination of the \Ni\ heating and the magnetar spin-down heating is considered for the double peaked Type~Ib SN~2005bf \citep[][]{maeda2007}. The combination of the CSM interaction and magnetar spin-down heating is also considered to explain multiple peaks in SLSN LCs \citep[e.g.,][]{chatzopoulos2016,li2020}.

While it is possible that several heating mechanisms play a role at the same time, it is also possible that multiple luminosity peaks are caused by a variation in a single energy source. It has been shown that luminosity variations can be caused if there are several shell structures in dense CSM \citep[][]{liu2018}. Whenever SN ejecta hit a shell, a temporal increase in luminosity causing a bumpy LC occurs. As we discuss in this paper, it is also possible that central energy injection from magnetar spin down is variable. However, the magnetar spin-down energy release is usually assumed to be a smooth decay and few studies consider the effect of variable magnetar spin-down energy inputs on LCs of stripped-envelope SNe \citep[e.g.,][]{kasen2016,chugai2021}. In this paper, we investigate the effects of variable magnetar spin-down energy injection on photometric properties of stripped-envelope SNe to identify their expected observational signatures.

The rest of this paper is organized as follows. We first introduce the thermal energy inputs from magnetar spin down and possible physical mechanisms causing their variabilities in Section~\ref{sec:variablemag}. In Section~\ref{sec:modeling}, we perform numerical LC modeling of SLSNe with variable thermal energy injection from magnetar spin down in order to clarify their effects on photometirc properties. We compare our models with SNe having bumpy LCs to discuss if the bumy LCs can be explained by variable energy injection in Section~\ref{sec:comparison}. The conclusion of this paper is summarized in Section~\ref{sec:conclusions}.

\section{Variable thermal energy injection from magnetar spin down}\label{sec:variablemag}
We assume that the spin down of magnetars powering SNe is mainly caused by a vaccum dipole magnetic field\footnote{Our discussion and conclusions do not change even if we assume a force-free magnetic field \citep[][]{contopoulos1999,spitkovsky2006,gruzinov2007}, although the estimated dipole magnetic field and initial spin period become different.} as often assumed in SN modeling with magnetar spin down. Then, the spin-down rate ($\dot{E}_\mathrm{rot}$) of magnetars is expressed as
\begin{equation}
    \dot{E}_\mathrm{rot} (t) = \frac{E_\mathrm{rot,0}}{t_\mathrm{sd}}\left(1+\frac{t}{t_\mathrm{sd}}\right)^{-2}, \label{eq:doterot}
\end{equation}
where $t$ is time, $E_\mathrm{rot,0}$ is the initial rotational energy of a magnetar, and $t_\mathrm{sd}$ is the spin-down time. Given a dipole magnetic field ($B_p$) and an initial spin period ($P_0$) of the magnetar, $E_\mathrm{rot,0}$ and $t_\mathrm{sd}$ can be expressed as
\begin{equation}
    E_\mathrm{rot,0} = \frac{1}{2}I \left(\frac{2\pi}{P_0}\right)^2 \simeq 2\times 10^{52}\left(\frac{P_0}{1~\mathrm{ms}}\right)^{-2}~\mathrm{erg},
\end{equation}
and
\begin{equation}
    t_\mathrm{sd} = \frac{3 I c^3 P_0^2}{4\pi^2 B_p^2 R^6 \sin^2\alpha} \simeq 4\times 10^{5} \left(\frac{P_0}{1~\mathrm{ms}}\right)^2\left(\frac{B_p}{10^{14}~\mathrm{G}}\right)^{-2}~\mathrm{sec}, \label{eq:tsd}
\end{equation}
respectively, where $I\simeq 10^{45}~\mathrm{g~cm^2}$ is the magnetar moment of inertia, $c$ is the speed of light, $R\simeq 10~\mathrm{km}$ is the magnetar radius, and $\alpha$ is the angle between the magnetic and rotation axes. We set $\alpha = 45^\circ$ in Eq.~(\ref{eq:tsd}).

A fraction of spin-down energy is thermalized to make SNe bright. This fraction is, however, uncertain. Many SLSN studies assume that all the spin-down energy is thermalized to power SLSNe. However, this is not justified in light of physics of pulsar wind nebulae. Most of the spin-down energy should be extracted as Poynting-dominated outflows \citep[e.g.,][]{goldreich1969}. The observations of Galactic pulsar wind nebulae suggest that a significant fraction must be dissipated around the termination shock as non-thermal energy \citep[e.g.,][]{kennel1984a,kennel1984b}. Thus, other studies assume that the spin-down energy is released as non-thermal electrons and calculate photon spectra, taking their opacity into account to estimate the fraction of high-energy photons to be thermalized \citep[e.g.,][]{kotera2013,metzger2014,murase2015,wang2015,kashiyama2016,nicholl2017,vurm2021}. 
In this study, we adopt a simplified parametarization to convert the spin-down energy to thermal energy powering SN LCs. Namely, we simply assume that a fraction $\varepsilon_\mathrm{th}$ of spin-down energy is thermalized and the thermal energy input from magnetar spin down is expressed as
\begin{equation}
    L_\mathrm{mag}(t)=\varepsilon_\mathrm{th}(t) \dot{E}_\mathrm{rot}(t). \label{eq:Lmag}
\end{equation}
The aim of this paper is to investigate the effect of variable thermal energy injection from magnetar spin down. The effect of variable thermal energy injection is taken into account by changing $\varepsilon_\mathrm{th}(t)$ in the following sections. Several mechanisms that may temporarily change $\varepsilon_\mathrm{th}(t)$ can be considered.

The thermalization efficiency $\varepsilon_\mathrm{th}(t)$ can be interpreted to have two components, i.e., $\varepsilon_\mathrm{th}(t)=\varepsilon_\mathrm{e}(t) f_\mathrm{dep,\gamma}(t)$. The first component $\varepsilon_\mathrm{e}(t)$ is the conversion efficiency from the spin-down energy to electron-positron pairs that form pulsar winds. 
The electron-positron pairs are further accelerated to high energies and release high-energy photons via synchrotron and inverse-Compton emission mechanisms. The gamma rays will be absorbed in the SN ejecta at early times. The second component $f_\mathrm{dep,\gamma}(t)$ is the fraction of high-energy photons that will be absorbed and thermalized in the SN ejecta. The factor $f_\mathrm{dep,\gamma}(t)$ can be regarded as an effect of opacity.

The conversion efficiency from spin-down energy to the nonthermal particle energy, $\varepsilon_\mathrm{e}(t)$, is considered to be close to unity in Galactic pulsar wind nebulae \citep[e.g.,][]{kennel1984a,kennel1984b,tanaka2011}. However, for newborn magnetars with the magnetization parameter that is much larger than unity, most spin-down energy may remain as the magnetic energy at the termination shock \citep[e.g.,][]{arons2012,murase2014}. In such a case, $\varepsilon_\mathrm{e}(t)$ may be much lower than unity, but could temporarily increase due to magnetic dissipation in the nebula \citep[e.g.,][]{porth2014}.

The absorption fraction $f_\mathrm{dep,\gamma}(t)$ can also change with time. For example, pair multiplicity injected into magnetar winds might be temporarily increased by, e.g., a flaring activity of magnetars. The temporal increase in pair multiplicity leads to the temporal reduction in injected particle energy and, therefore, the temporal energy reduction in photons radiated from high-energy particles. Because opacity of high-energy photons strongly depends on photon energy \citep[e.g.,][]{kotera2013,murase2015,murase2021,badjin2016,vurm2021}, the temporal change in the energy of injected photons can lead to a temporal change in $f_\mathrm{dep,\gamma}(t)$. The factor $f_\mathrm{dep,\gamma}(t)$ may increase or declease depending on the original spectral energy distributions.

Effects of variable thermal energy injection from magnetar spin down due to inefficient thermalization by the reduction in $\varepsilon_\mathrm{e}(t)$ and $f_\mathrm{dep,\gamma}(t)$ was previously studied by \citet{kasen2016}. They considered the effect of inefficient thermalization shortly after the magnetar formation in order to explain precursors sometimes observed in SLSNe \citep[][]{nicholl2016bump,angus2019}. In this study, we focus on variable thermal energy injection at later phases, mostly after the main luminosity peak.

In this work, we change $\varepsilon_\mathrm{th}(t)$ to investigate the effect of variable thermal energy injection from magnetar spin-down. However, we can also consider other ways to change the spin-down energy injection rate with time. For example, the variable magnetar energy injection can be realized by changing the angle between the magnetic and rotation axes ($\alpha$). The angle determines the spin-down time (Eq.~\ref{eq:tsd}) and a temporal change in the angle can result in a temporal variation in the magnetar energy injection rate. The angle could naturally get larger because having a larger angle leads to a lower moment of inertia and, therefore, a lower energy state. In this case, the thermalization rate does not need to change to have a variable energy injection.

Another similar mechanism to cause the variable magnetar spin-down energy injection is to make a variation in the magnetar spin. The fallback accretion towards the central magnetar can provide angular momentum that accelerates the rotation of the magnetar (e.g., \citealt{barkov2011}; \citealt{lin2021}). The spin-up of the magnetar results in the increase of the energy injection. This variation is, again, not a simple increase in $\varepsilon_\mathrm{th}$. If the fallback accretion occurs intermittently, the energy injection variation can be observed several times.

\section{Effect of variable thermal energy injection on SLSN properties}\label{sec:modeling}
We investigate the effects of the variable thermal energy injection from magnetar spin down discussed so far on observational properties of SLSNe in this section.

\subsection{Methods}
In order to quantify the effect of the variable magnetar energy injection on SLSN observational properties, we perform numerical LC calculations of SLSNe powered by magnetar spin down. We use one-dimensional multi-frequency radiation hydrodynamics code \texttt{STELLA} \citep[][]{blinnikov1998,blinnikov2000,blinnikov2006} for this purpose. Briefly, \texttt{STELLA} implicitly calculates time-dependent equations of the angular moments of intensity averaged over a frequency bin using the variable Eddington method. In a standard setup, 100 frequency bins are set from 1~\AA\ to 50000~\AA\ in log scale. By convolving filter response functions to calculated spectral energy distributions (SEDs), we can estimate multi-band LCs. We define photosphere at the location where the Rosseland-mean optical depth becomes $2/3$ at each timestep in order to estimate photospheric properties. \texttt{STELLA} was previously used to obtain LCs powered by magnetar spin down \citep[][]{moriya2016mag} and we adopt a similar method in this paper. We refer to \citet{blinnikov1998,blinnikov2000,blinnikov2006} for the further details of \texttt{STELLA}.

We put SN progenitor models described below as the initial condition in \texttt{STELLA}. The mass cut, below which is assumed to form a neutron star after core collapse, is set at 1.4~\Msun. In other words, the progenitor structure above 1.4~\Msun\ is put into \texttt{STELLA} and the central 1.4~\Msun\ is treated as a central point source. We first initiate SN explosions by putting thermal energy just above the mass cut for 0.1~sec. The injected thermal energy at this first 0.1~sec is set to achieve the SN explosion energy of $5\times 10^{51}~\mathrm{erg}$, which is a typical value in SLSNe \citep[e.g.,][]{nicholl2015,nicholl2017,blanchard2020}. We do not follow explosive nucleosynthesis, and the luminosity input from the radioactive decay of \Ni\ is assumed to be negligible.

We start to inject the magnetar spin-down energy from 0.1~sec. $L_\mathrm{mag}(t)$ (Eq.~\ref{eq:Lmag}) is injected as thermal energy just above the mass cut. The temperature of the central region in which the thermal energy is injected is determined by $(E_\mathrm{in}/4\pi a R_\mathrm{in}^3)^{1/4}$, where $E_\mathrm{in}$ is the injected thermal energy, $R_\mathrm{in}$ is the radius of the energy injection, and $a$ is the radiation constant, because the radiation pressure is dominant. We set $\varepsilon_\mathrm{th}(t) = 0.1$ as a default value and temporarily change $\varepsilon_\mathrm{th}$ to see the effect of variable magnetar energy injection. We note that both $\varepsilon_e$ and $f_\mathrm{dep,\gamma}$ are likely close to unity shortly after explosion (e.g., \citealt{kashiyama2016,omand2018}, but see also \citealt{kasen2016}) and then $f_\mathrm{dep,\gamma}$ can get smaller to make $\varepsilon_\mathrm{th}$ smaller as the ejecta expand \citep[e.g.,][]{murase2021}. In this work, we artificially set the default value of $\varepsilon_\mathrm{th}$ to be small (0.1) in order to show the effects of variable $\varepsilon_\mathrm{th}$ including the possible temporal increase of $\varepsilon_\mathrm{th}$ at any epochs.

We adopt two SN progenitor models presented in \citet{sukhbold2016}. The first one (s70) has 70~\Msun\ at zero-age main sequence and becomes a 6.4~\Msun\ Wolf-Rayet star without hydrogen and helim at core collapse. If we assume the mass cut at 1.4~\Msun, the ejecta mass from explosions of this progenitor becomes 5~\Msun, which is a typical ejecta mass for SLSNe \citep[e.g.,][]{nicholl2015,nicholl2017,blanchard2020}. The other progenitor model we use is s45, which has the zero-age main-sequence mass of 45~\Msun\ and evolves to a 13~\Msun\ Wolf-Rayet star without hydrogen and helium at the time of core collapse. The ejecta mass from this progenitor is 11.6~\Msun\ with the mass cut at 1.4~\Msun. This progenitor model is used in Section~\ref{sec:comparison} where we investigate the origins of the LC bumps observed in SN~2015bn and SN~2019stc.

The progenitor models from \citet{sukhbold2016} are calculated as a non-rotating single star with the solar metallicity. The hydrogen-rich and helium layers of the progenitors are stripped through the strong stellar wind. However, rotation is essential in powering SLSNe with magnetar spin down, and SLSN progenitors could evolve in binary system \citep[e.g.,][]{aguilera-dena2018}. In addition, SLSNe tend to appear in low metallicity environments \citep[e.g.,][]{schulze2018,wiseman2020}. Even though the exact evolutionary path can affect the structure of SN progenitors \citep[e.g.,][]{schneider2021,laplace2021}, the SN photometric properties are mainly determined by the ejecta mass and explosion energy. The slight difference in metallicity does not affect the SN photometric properties, either. Thus, the progenitor models we use are good enough to investigate the effect of variable thermal energy injection from magnetar spin down.

\begin{figure}
	\includegraphics[width=\columnwidth]{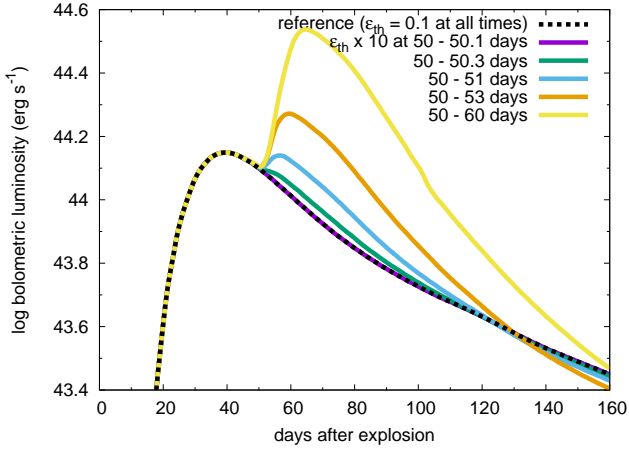}	
    \caption{
    Bolometric LCs with temporal magnetar thermal energy injection increase by a factor of 10 starting at 50~days, which is 10~days after the bolometirc LC peak of the reference model (dashed line) with the constant $\varepsilon_\mathrm{th}=0.1$ at all times.
    }
    \label{fig:tn50_lc}
\end{figure}

\begin{figure}
	\includegraphics[width=\columnwidth]{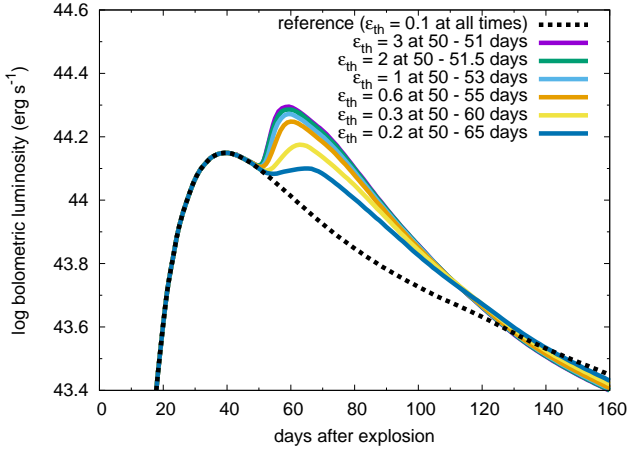}
    \caption{
    Bolometric LCs with temporal magnetar thermal energy injection increase starting at 50~days with the same $\varepsilon_\mathrm{th}\Delta t = 3$. The models with $\varepsilon_\mathrm{th}>1$ are shown for demonstration. The reference model (dashed line) has the constant $\varepsilon_\mathrm{th}=0.1$ at all times.
    }
    \label{fig:tn50_epsdtconst}
\end{figure}

\begin{figure}
	\includegraphics[width=0.95\columnwidth]{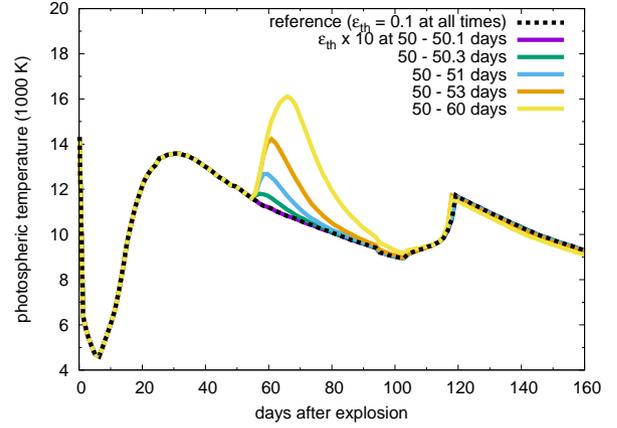}
	\includegraphics[width=0.95\columnwidth]{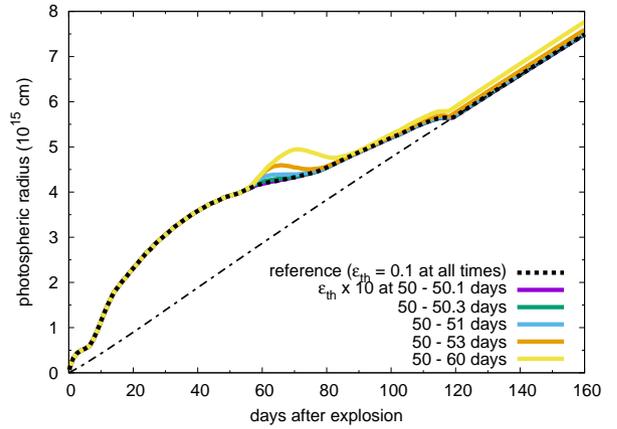}		
	\includegraphics[width=0.95\columnwidth]{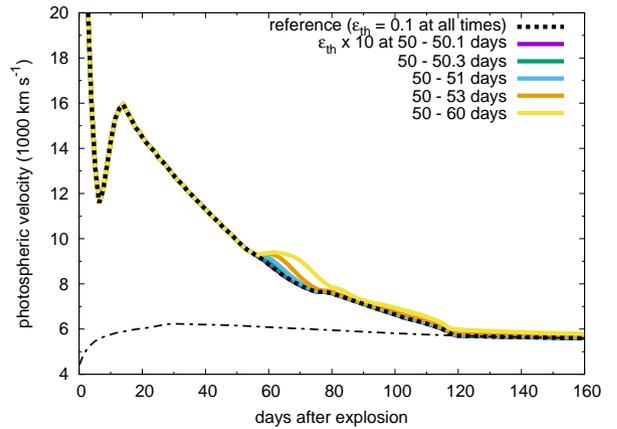}	
    \caption{
    Photospheric temperature (top), radius (middle), and velocity (bottom) evolution of the models in which the magnetar energy conversion efficiency, which is set as  $\varepsilon_\mathrm{th}=0.1$ at all times in the reference model (dashed lines), is temporarily increased by a factor of 10 ($\varepsilon_\mathrm{th}=1$). The bolometric LCs of the models are presented in Fig.~\ref{fig:tn50_lc}. The dot-dashed lines in the middle and bottom panels show the location of the innermost layer in our numerical calculations that corresponds to the termination shock.
    }
    \label{fig:tn50_ph}
\end{figure}

\subsection{Results}
\subsubsection{Reference model}
We first introduce our reference model in which we do not include variability in the magnetar energy injectioin, i.e., $\varepsilon_\mathrm{th} = 0.1$ at all time in Eq.~(\ref{eq:Lmag}). We adopt the 5~\Msun\ ejecta model and the explosion energy is set to $5\times 10^{51}~\mathrm{erg}$ in all the models. Assuming the electron scattering opacity of $0.2~\mathrm{cm^2~g^{-1}}$, the corresponding diffusion time of the ejecta is 32~days.

We set $B_p=3\times 10^{13}~\mathrm{G}$ and $P_0=1~\mathrm{ms}$, which corresponds to $E_\mathrm{rot,0}=2\times 10^{52}~\mathrm{erg}$ and $t_\mathrm{sd}=52~\mathrm{days}$. The bolometric LC from the reference model is presented in Fig.~\ref{fig:tn50_lc} with a dashed line. It has the rise time of 40~days and the peak bolometric luminosity of $1.4\times 10^{44}~\mathrm{erg~s^{-1}}$. Photospheric properties of the reference model can be found in Fig.~\ref{fig:tn50_ph}. We introduce variable thermal energy injection to this model and discuss its effect compared to this reference model from next section.

\begin{figure}
	\includegraphics[width=\columnwidth]{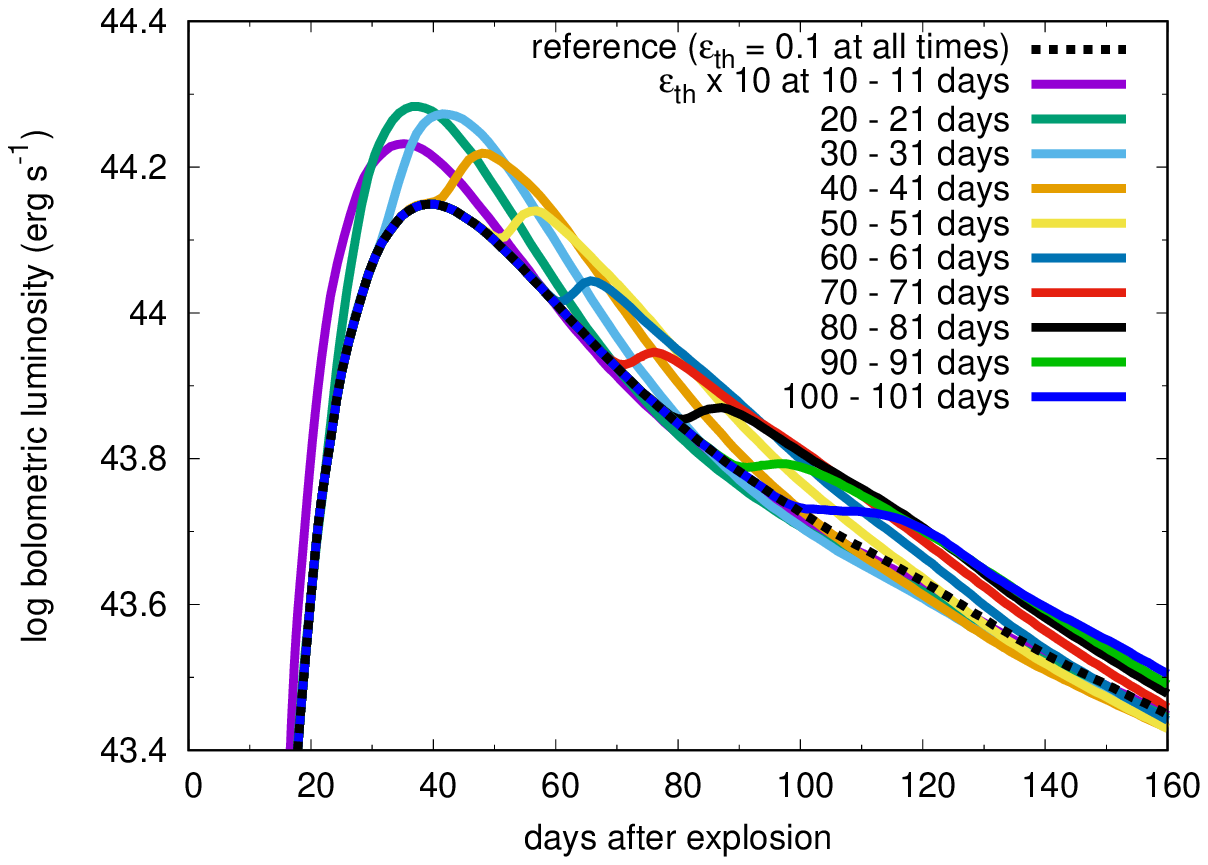}
	\includegraphics[width=\columnwidth]{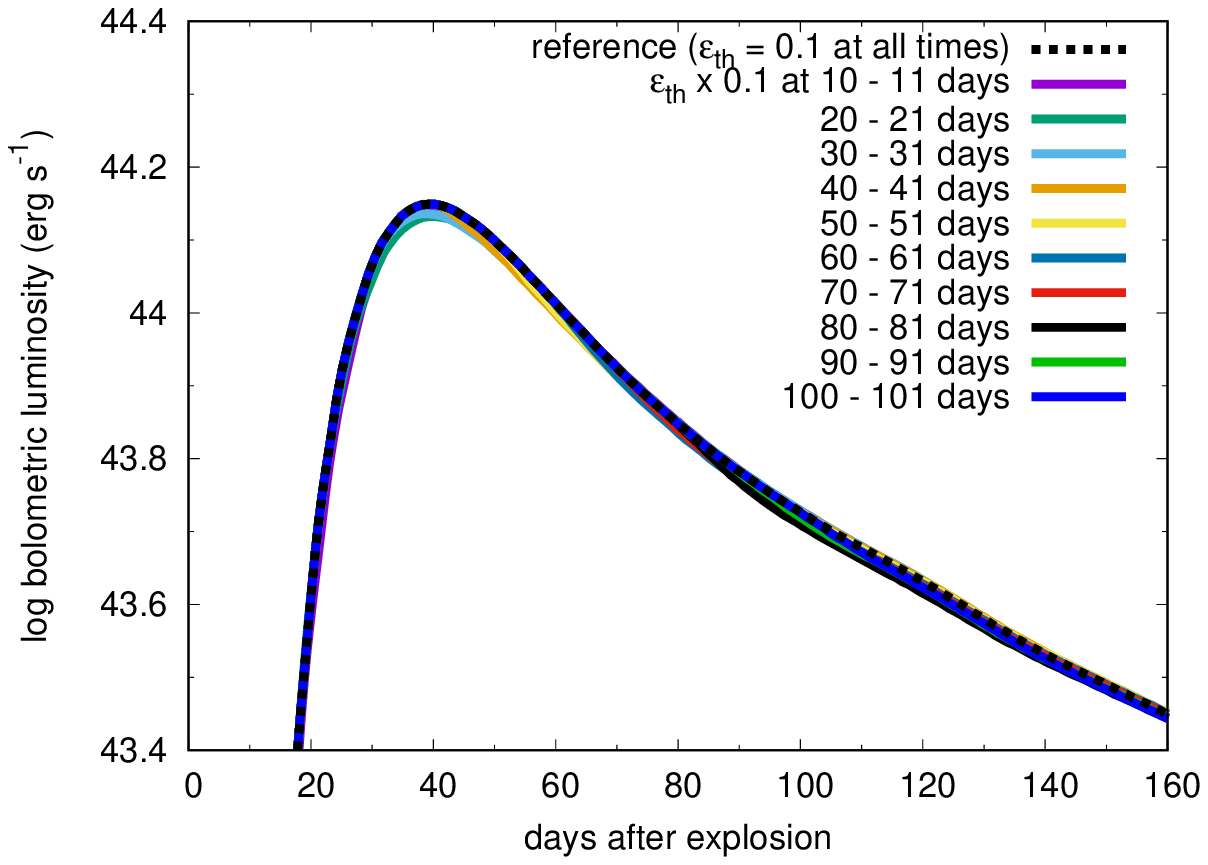}	
    \caption{
    Bolometric LCs in which the magnetar energy injection efficiency $\varepsilon_\mathrm{th}$ is temporarily changed at various times. The top panel shows the cases of the temporal energy injection efficiency increase from the original $\varepsilon_\mathrm{th}=0.1$ to $\varepsilon_\mathrm{th}=1$. The bottom panel shows the cases of the temporal energy injection efficiency decrease from the original $\varepsilon_\mathrm{th}=0.1$ to $\varepsilon_\mathrm{th}=0.01$.
    }
    \label{fig:tnx10_tnx0p1}
\end{figure}

\begin{figure}
	\includegraphics[width=\columnwidth]{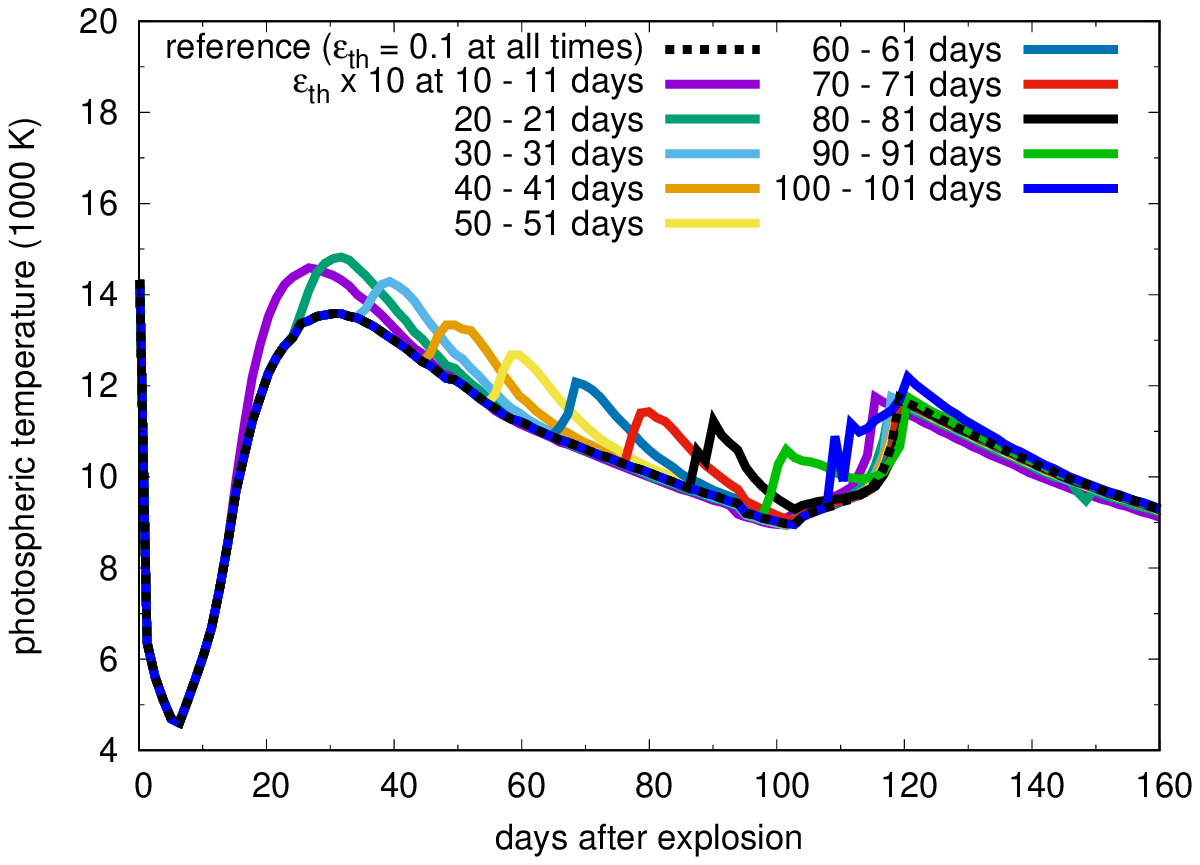}
	\includegraphics[width=\columnwidth]{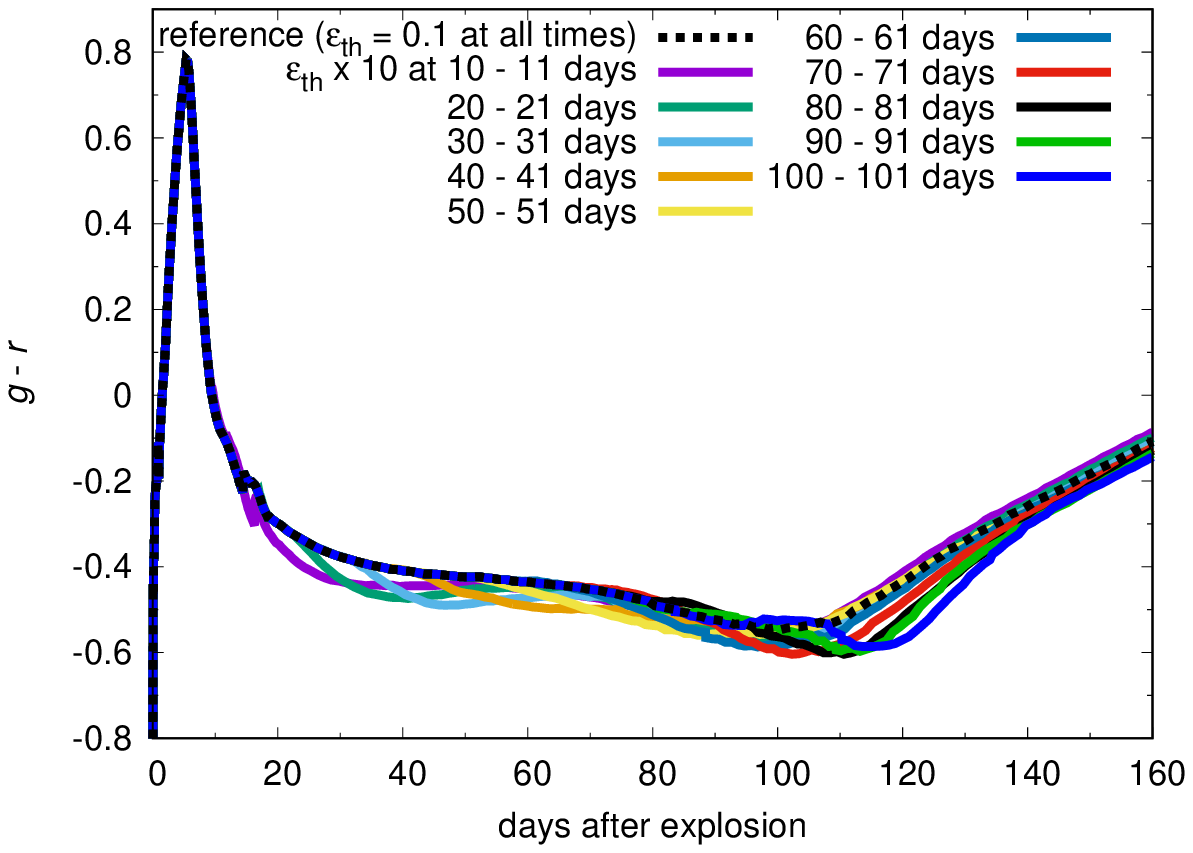}	
    \caption{
    Photospheric temperature (top) and $g-r$ color (bottom) evolution of the models in which the magnetar energy injection efficiency is temporarily increased from $\varepsilon_\mathrm{th}=0.1$ to 1. The corresponding bolometric LCs are presented in the top panel of Fig.~\ref{fig:tnx10_tnx0p1}.
    }
    \label{fig:temp_col_tnx10}
\end{figure}

\subsubsection{Temporal energy increase after the LC peak}\label{sec:afterpeak}
We first temporarily increase $\varepsilon_\mathrm{th}$ from 50~days, which is 10~days after the bolometric LC peak. Fig.~\ref{fig:tn50_lc} shows the changes in the bolometric LCs when we increase $\varepsilon_\mathrm{th}$ by a factor of 10 at $50-50.1~\mathrm{days}$ ($\Delta t = 0.1~\mathrm{days}$), $50-50.3~\mathrm{days}$ ($\Delta t = 0.3~\mathrm{days}$), $50-51~\mathrm{days}$ ($\Delta t = 1~\mathrm{day}$), $50-53~\mathrm{days}$ ($\Delta t = 3~\mathrm{days}$), and $50-60~\mathrm{days}$ ($\Delta t = 10~\mathrm{days}$). Here, $\Delta t$ is the duration of the temporal $\varepsilon_\mathrm{th}$ change. We start to see the clear second bump caused by the increased energy injection if the energy increase is kept for more than 1~day.

Fig.~\ref{fig:tn50_epsdtconst} shows the bolometric LCs with the temporal $\varepsilon_\mathrm{th}$ increase from 50~days with the constant $\varepsilon_\mathrm{th}\Delta t = 3$. We can find that the effect of the temporal $\varepsilon_\mathrm{th}$ increase in these models is the same when $\Delta t$ is so small that the diffusion in the ejecta determines the subsequent bolometric LC properties. Then, the peak time of the second bump increases as $\Delta t$ increases. 

Fig.~\ref{fig:tn50_ph} shows the photospheric temperature ($T_\mathrm{ph}$), radius ($R_\mathrm{ph}$), and velocity of the models in which $\varepsilon_\mathrm{th}$ is temporarily increased by a factor of 10. The dot-dashed lines in the panels of the photospheric radius and velocity show the location of the innermost layer where thermal energy from the magnetar spin-down is injected. It corresponds to the termination shock in our calculations. In the reference model, the ejecta cool adiabatically after the shock breakout and the photospheric temperature keeps decreasing. Then, the photospheric temperature increases from around 7~days as the thermal energy injection from the magnetar spin down heats the ejecta. The SN ejecta become optically thin when the dot-dashed line matches the photospheric radius in Fig.~\ref{fig:tn50_ph} which occurs at 120~days after the explosion. At this moment, the photosphere no longer exists and the photospheric information shown in the figure is unphysical. Especially, the photospheric temperature increase from 100~days to 120~days is an artifact caused by the proximity of the photosphere and the location of the thermal energy injection.

Comparing the photosphere information in Fig.~\ref{fig:tn50_ph} with the bolometric LCs presented in Fig.~\ref{fig:tn50_lc}, we can find that there is significant increase in photospheric temperature during the temporal $\varepsilon_\mathrm{th}$ increase, while changes in photospheric radii and, therefore, velocity, are minor. This shows that the temperature increase is the major cause of the bolometric luminosity increase ($4\pi R^2 \sigma T^4$, where $\sigma$ is the Stefan--Boltzmann constant) when a second LC peak appears due to the temporal increase in the central thermal energy injection. The temporal increase of the injected thermal energy leads to the temperature increase of the ejecta. However, the inner temperature growth is not large enough to change the ionization state of the ejecta at around the photosphere. Thus, the photospheric radius is not strongly altered by the inner temperature growth and the photospheric temperature increase is mainly observed.

\begin{figure}
	\includegraphics[width=\columnwidth]{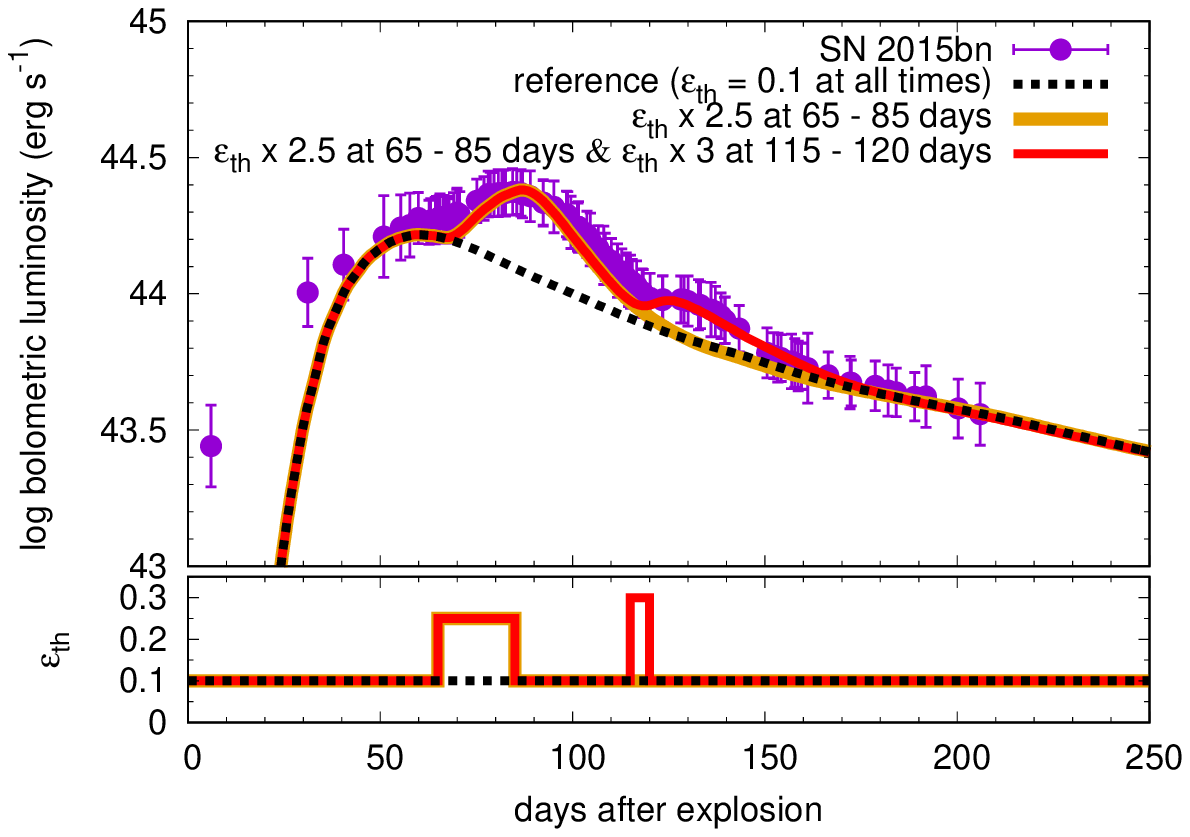}
	\includegraphics[width=\columnwidth]{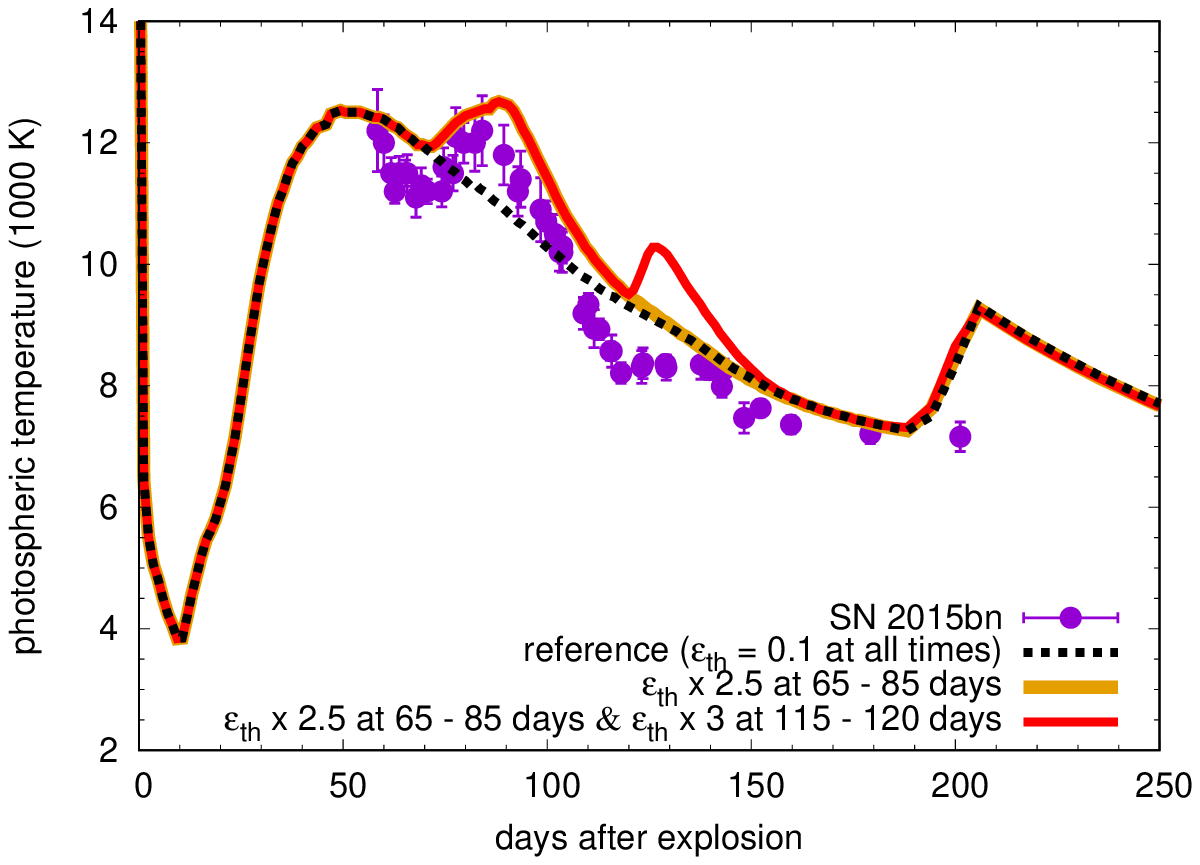}	
	\includegraphics[width=\columnwidth]{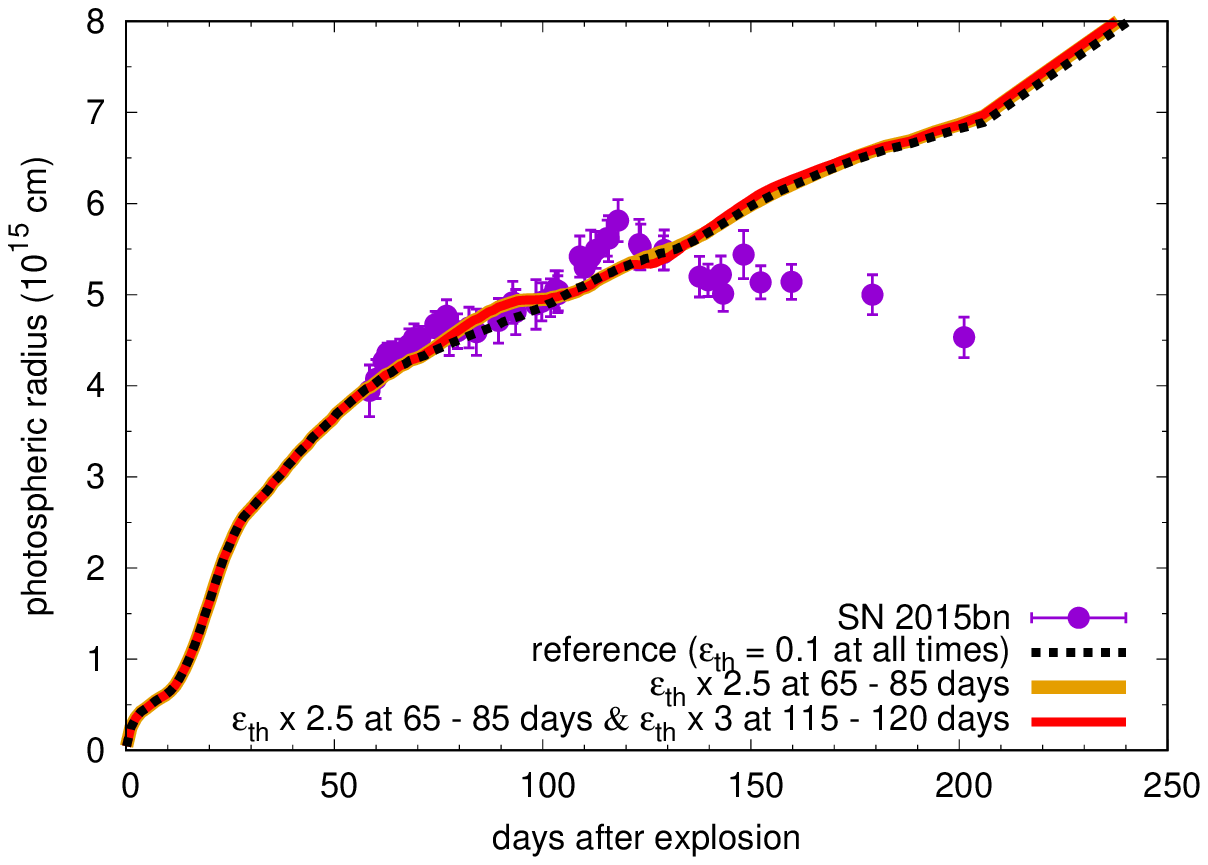}
    \caption{
    Bolometric LC (top), photospheric temperature (middle), and photospheric radius of SN~2015bn and their comparison with our synthetic models with temporal $\varepsilon_\mathrm{th}$ variations. The top panel also shows $\varepsilon_\mathrm{th}(t)$. The observational data are taken from \citet{nicholl2016sn2015bn1}. The photospheric radii keep increasing in the synthetic models even after the ejecta become optically thin, which is beyond the applicability of the models (Section~\ref{sec:afterpeak}).
    }
    \label{fig:sn2015bn}
\end{figure}

\subsubsection{Temporal energy variations at different epochs}
In the previous section, we presented the models in which $\varepsilon_\mathrm{th}$ is temporarily increased after the bolometric LC peak of the reference model. Here we show the models in which $\varepsilon_\mathrm{th}$ is temporarily changed at various times.

Fig.~\ref{fig:tnx10_tnx0p1} shows the bolometric LC models in which $\varepsilon_\mathrm{th}$ is temporarily changed from the original value of $\varepsilon_\mathrm{th}=0.1$ at various times. First, the bolometric LCs in the top panel of Fig.~\ref{fig:tnx10_tnx0p1} are those with the temporal magnetar spin-down energy input increase by a factor of 10. We can find that the temporal energy increase before the bolometric LC peak at 40~days does not lead to multiple LC peaks. The rise time of bolometric LCs corresponds to the diffusion time of the ejecta. The thermal energy added before the bolometric LC peak simply diffuses out in the original diffusion timescale without making an additional peak in the bolometric LCs. When the energy increase occurs after the bolometric LC peak, a secondary bolometric LC peak can be observed. As discussed in the previous section, photospheric temperature is mainly affected by the additional energy input. Fig.~\ref{fig:temp_col_tnx10} shows the photospheric temperature and $g-r$ color evolution of the models presented in the top panel of Fig.~\ref{fig:tnx10_tnx0p1}. 
We can find that the increase in photospheric temperature is accompanied by increase in bolometric luminosity. The $g-r$ color becomes bluer by about 0.1~mag during the increase of luminosity.

The bottom panel of Fig.~\ref{fig:tnx10_tnx0p1} shows the bolometric LC models in which the magnetar energy injection is temporarily reduced by a factor of 10. We do not find significant effects on the photometric properties when we reduce the energy injection temporarily.

\section{Comparisons with observations}\label{sec:comparison}
We have shown the effects of temporal variations in injected thermal energy in magnetar-powered SN models. It is found that temporal injected energy increase after the bolometric LC peak can result in the temporal increase in luminosity. Photospheric temperature is significantly increased during the temporal luminosity increase, while the photospheric radius and velocity are less affected. In this section, we compare our models with the observational properties of luminous SNe with multiple luminosity peaks.

\subsection{SN 2015bn}
SN~2015bn is a SLSN known to have bumpy LCs as reported in \citet{nicholl2016sn2015bn1,nicholl2016sn2015bn2}. We take the bolometric LC and photospheric properties reported by \citet{nicholl2016sn2015bn1} and compare them with our model with variable thermal energy injection from magnetar spin down. We use the s45 progenitor model having the ejecta mass of 11.6~\Msun\ for this purpose. The progenitor model is exploded to have the explosion energy of $5\times 10^{51}~\mathrm{erg}$ with a thermal bomb. Our reference model with the constant $\varepsilon_\mathrm{th}=0.1$ has $B_p = 2\times 10^{13}~\mathrm{G}$ and $P_0=0.8~\mathrm{ms}$, which corresponds to $E_\mathrm{rot,0}=3.1\times 10^{52}~\mathrm{erg}$ and $t_\mathrm{sd}=75~\mathrm{days}$. Other studies obtain different $B_p$ and $P_0$ because of different assumptions on $\varepsilon_\mathrm{th}$ and the spin-down formula \citep[e.g.,][]{nicholl2016sn2015bn1,omand2018}. The reference model has the rise time of 61~days and the peak bolometric luminosity of $1.6\times 10^{44}~\mathrm{erg~s^{-1}}$. We temporarily change $\varepsilon_\mathrm{th}$ from the reference model to match the observed properties.

Fig.~\ref{fig:sn2015bn} compares the observed bolometric LC, photospheric temperature, and photospheric radius with our model. When we construct the bolometric LC models to compare, we ignore the first observation at 6~days because it may have been a precursor often found in SLSNe \citep{nicholl2016sn2015bn1}. We succeed in reproducing the two bumps observed in the bolometric LCs by introducing two phases of the temporal increase of $\varepsilon_\mathrm{th}$ \citep[see also][]{yu2017}. We first increase $\varepsilon_\mathrm{th}$ by a factor of 2.5 from $65~\mathrm{days}$ to $85~\mathrm{days}$ to match the major bolometric LC peak at around 90~days. Then, increasing $\varepsilon_\mathrm{th}$ by a factor of 3 from 115~days to 120~days enables us to account for the luminosity increase starting at 120~days. The extra thermal energy released during the temporal increase in $\varepsilon_\mathrm{th}$ is $3.1\times 10^{50}~\mathrm{erg}$ in $65-85~\mathrm{days}$ and $6.3\times 10^{49}~\mathrm{erg}$ in $115-120~\mathrm{days}$.

The luminosity increase in the synthetic model caused by the temporal increase in $\varepsilon_\mathrm{th}$ is clearly accompanied by photospheric temperature increase as discussed in Section~\ref{sec:modeling}. The photospheric temperature of SN~2015bn estimated from the observations around the bolometric LC peak indeed show clear temporal increase in photospheric temperature (Fig.~\ref{fig:sn2015bn}). On the other hand, the second LC bump starting from around 120~days does not have a clear photospheric temperature increase as in the previous LC peak. The observed photospheric radius has a slight increase during this bump which is not found in the synthetic model. Thus, the first bump causing the luminosity peak at around 90~days is consistent with the variable thermal energy injection from magnetar spin down, while it is difficult to conclude that the second bump is caused by the variable thermal energy injection. 

\begin{figure}
	\includegraphics[width=\columnwidth]{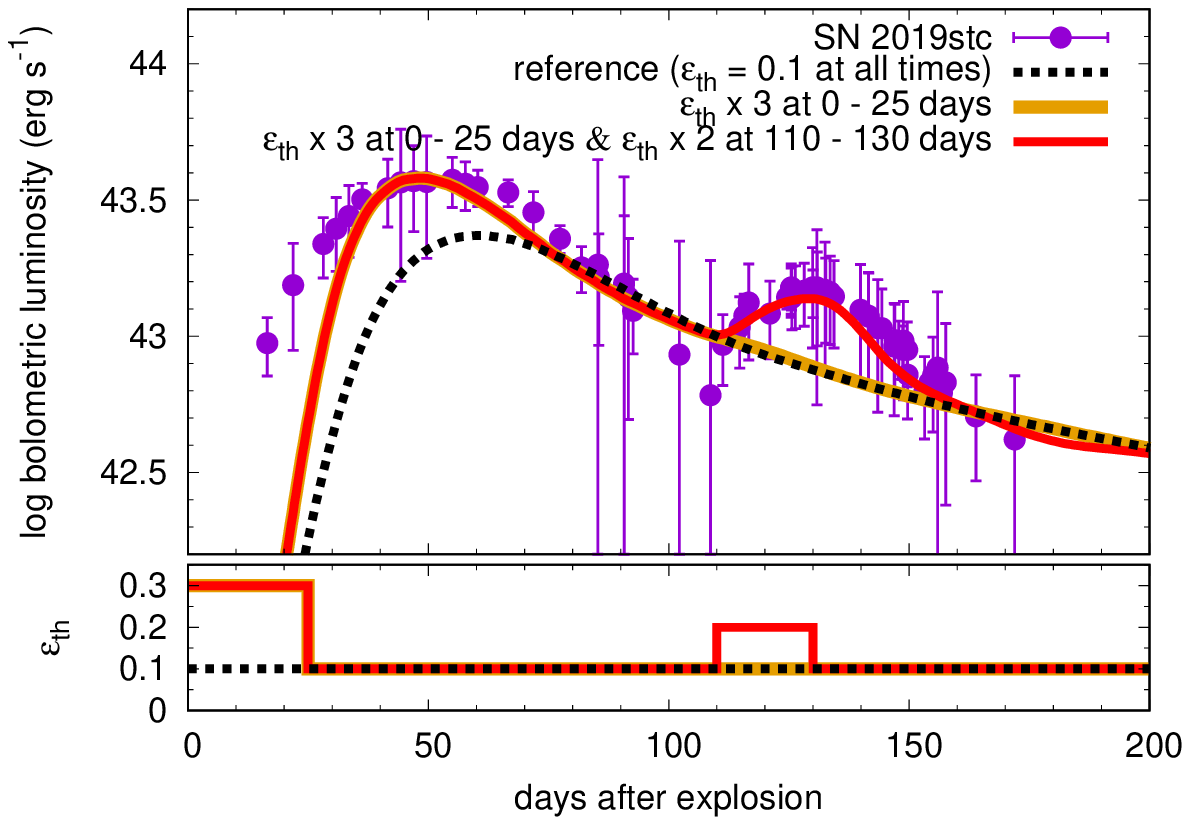}
	\includegraphics[width=\columnwidth]{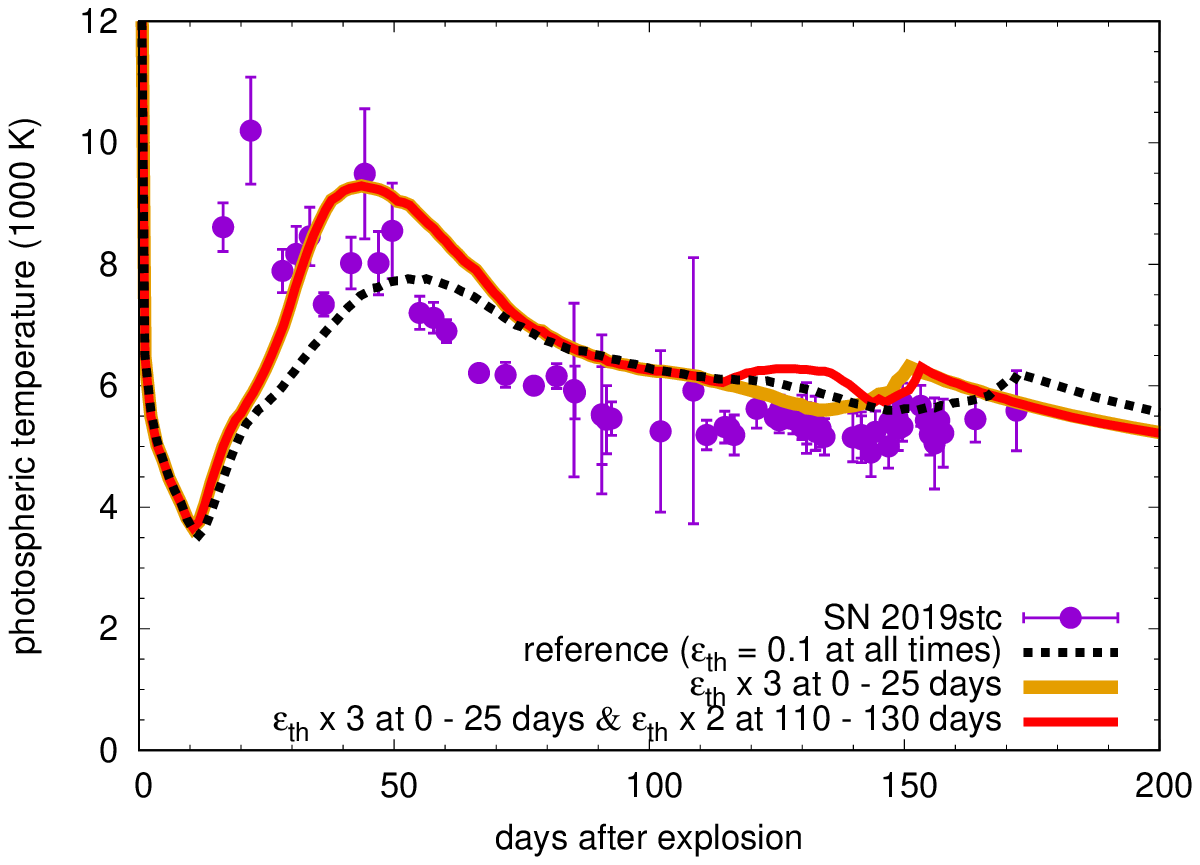}
	\includegraphics[width=\columnwidth]{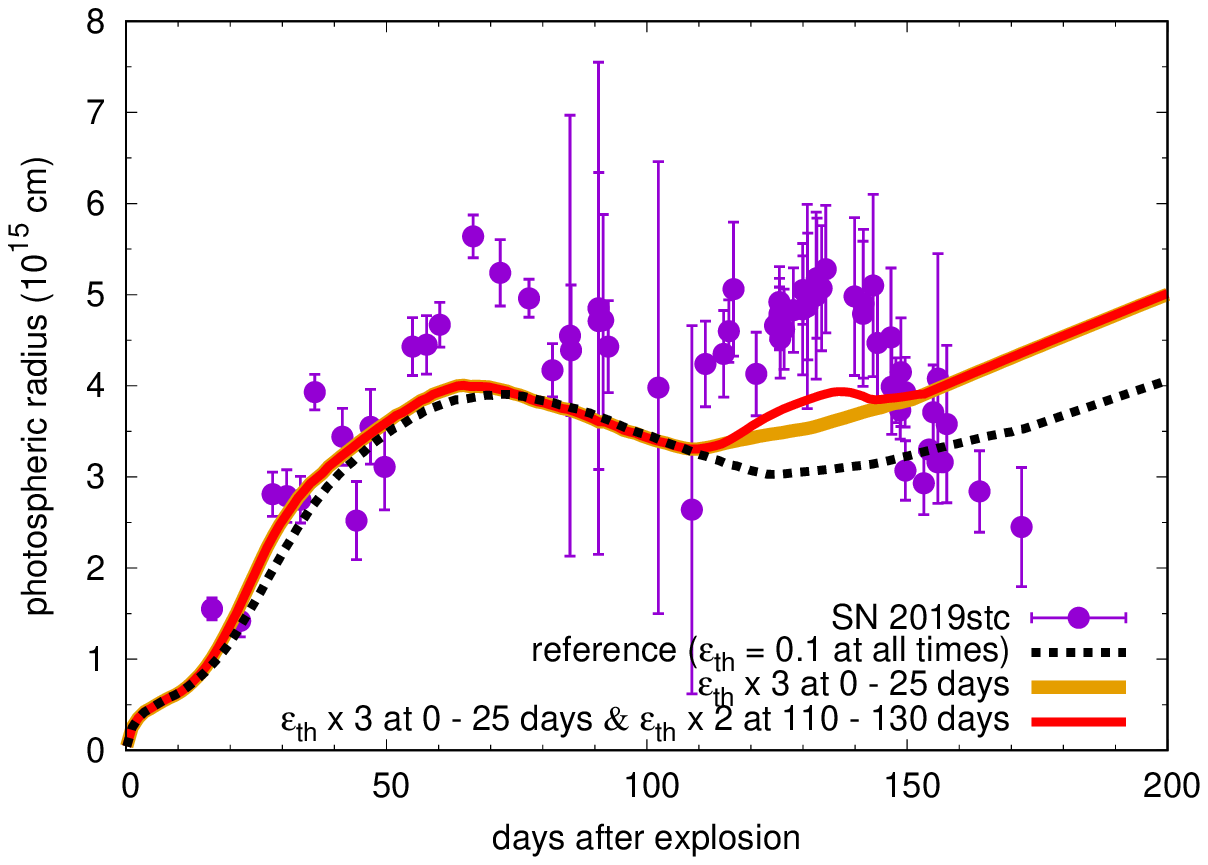}	
    \caption{
    Bolometric LC (top), photospheric temperature (middle), and photospheric radius of SN~2019stc and their comparison with our synthetic models with temporal $\varepsilon_\mathrm{th}$ variations. The top panel also shows $\varepsilon_\mathrm{th}(t)$. The observational data are taken from \citet{gomez2021}.
    }
    \label{fig:sn2019stc}
\end{figure}

\subsection{SN 2019stc}
SN~2019stc is a luminous Type~Ic SN with the peak bolometric luminosity of around $4\times 10^{43}~\mathrm{erg~s^{-1}}$ \citep{gomez2021}. It is not as luminous as typical SLSNe, but the peak luminosity is difficult to reach by the canonical \Ni\ heating, and it is suggested to be powered by magnetar spin down \citep{gomez2021}. The bolometric LC of SN~2019stc has a clear secondary peak as shown in Fig.~\ref{fig:sn2019stc}. We investigate if the secondary peak can be explained by the variable energy input from magnetar spin down. We adopt the same progenitor model as used for SN~2015bn in the previous section (s45 progenitor model). The explosion energy is set as $5\times 10^{51}~\mathrm{erg}$. Our models have $B_p = 7\times 10^{13}~\mathrm{B}$ and $P_0=2.4~\mathrm{ms}$. They correspond to $E_\mathrm{rot,0}=3.5\times 10^{51}~\mathrm{erg}$ and $t_\mathrm{sd}=55~\mathrm{days}$. The reference model has a constant $\varepsilon_\mathrm{th}=0.1$ and we temporarily vary $\varepsilon_\mathrm{th}$ to match the bolometric LC of SN~2019stc.

Fig.~\ref{fig:sn2019stc} shows the comparison between the observations and synthetic models. The first observed peak is matched by temporarily increasing $\varepsilon_\mathrm{th}$ by a factor of 3 from 0~days to 25~days. As presented in Section~\ref{sec:modeling}, temporal increase of $\varepsilon_\mathrm{th}$ starting before the bolometric LC peak of the reference model does not lead to multiple LC peaks but alters the first peak already seen in the reference model. The temporal $\varepsilon_\mathrm{th}$ increase starting before the bolometric LC peak allows us to match the observed first peak well. The bolometric luminosities in the first few epochs are still brighter than our synthetic LC. This may be because of the existence of a precursor \citep[e.g.,][]{nicholl2016bump}, a small ejecta mass, or a large explosion energy in our LC modeling. Then, the secondary bolometric LC peak is reproduced by increasing $\varepsilon_\mathrm{th}$ by a factor of 2 from 110~days to 130~days. The extra thermal energy released during the temporal increase in $\varepsilon_\mathrm{th}$ is $2.2\times 10^{50}~\mathrm{erg}$ at $0-25~\mathrm{days}$ and $2.5\times 10^{49}~\mathrm{erg}$ at $110-130~\mathrm{days}$. \citet{chugai2021} recently showed a similar model to increase the magnetar spin-down energy input to explain the secondary bolometric LC peak.

Interestingly, SN~2019stc does not show a clear increase in photospheric temperature during the secondary bolometric LC peak as shown in Fig.~\ref{fig:sn2019stc}. The observed bolometric luminosity increase is mostly due to the increase in the photospheric radius. However, our models with temporal increase in thermal energy injection from magnetar spin down predict no significant increase in the photospheric radius as shown in Fig.~\ref{fig:sn2019stc}. Therefore, we argue that the bumpy LC observed in SN~2019stc is not necessarily caused by variable thermal energy injection from magnetar spin down. \citet{chugai2021} concluded that the multiple LC peaks of SN~2019stc can be explained by increasing the magnetar energy input just by using LC information. Our results show that it is important to check photospheric properties to make constraints on the cause of the bumpy LCs.

\subsection{Implications}
We have compared the observational properties of SN~2015bn and SN~2019stc, which are luminous SNe with bumpy LCs, with our synthetic models with temporal variation in $\varepsilon_\mathrm{th}$ to see if the bumpy LCs can be explained by the variable thermal energy injection from magnetar spin down. We showed that the bumpy bolometric LCs of the two SNe can be reproduced by the temporal variability in the thermal energy injection by a factor of $2-3$. The required timescale for the temporal injected thermal energy increase is $\sim 1-10~\mathrm{days}$. This timescale can be comparable to the the thermalization timescale in the SN ejecta with the radius of around $10^{15}~\mathrm{cm}$, given the high-energy photon optical depth of the order of $10-100$. 
Because of the large required extra energy to make the LC bumps ($>10^{49}~\mathrm{erg}$), the temporal magnetic field energy release as observed in Galactic magnetars is not likely the cause of the LC bumps.

The bumpy LCs caused by the variable thermal energy injection are predicted to be accompanied by the significant photospheric temperature increase. While the first luminosity bump in SN~2015bn is observed to have the phtospheric temperature increase during the LC bumps as predicted by the variable magnetar energy injection, SN~2019stc does not have the photospheric temperature increase as predicted by our synthetic models. The bumpy LC of SN~2019stc is accompanied by the significant increase in photospheric radius which is not predicted in our models with variable thermal energy injection from magnetar spin down. We conclude that the variations in thermal energy injection do not explain all the bumpy LCs observed in luminous and superluminous SNe, and other mechanisms such as additional CSM interaction are also required to be considered. The LC bumps caused by the variable thermal energy injection from magnetars and those caused by the CSM interaction would lead to different photometric and spectroscopic properties. For example, the LC bumps from the CSM interaction could be accompanied with some narrow emission lines caused by the dense CSM. More studies on the LC bumps from the CSM interaction are required to identify the observational features that distinguish LC bumps from the thermalization variation and the CSM variation.

Finally, we note that our LC models from magnetar spin-down have simplifications. For example, even though the magnetar spin-down is a multi-dimensional process, our models assume a spherical symmetry. Jets may be released during magnetar spin-down \citep[e.g.,][]{soker2016} and they may be responsible for late-phase LC bumps \citep[e.g.,][]{kaplan2020}. Despite of our simplification, our one-dimensional models would provide a first-order quantitative estimate of the effects of the thermalization variation that would guide further detailed investigations.

\section{Conclusions}\label{sec:conclusions}
We have investigated the effect of variable thermal energy injection from magnetar spin down on photometric properties of luminous SNe including SLSNe. We assumed that thermalization efficiency of magnetar spin down energy can change temporarily and investigated their effects on the photometric properties. We showed that temporal increase in thermalizaion efficiency after the bolometric luminosity peak can result in bumpy LCs as observed in some luminous and superluminous SNe. We found that photospheric temperature is significantly increased during the LC bumps caused by the variable thermalization efficiency, while the photospheric radius is less affected.

We have found that the observed LC bumps in SN~2015bn and SN~2019stc can be reproduced by temporarily increasing the thermalization efficiency by a factor of $2-3$ for $5-20~\mathrm{days}$. However, some observed LC bumps, especially those in SN~2019stc, are not accompanied by the photospheric temperature increase as clear as our synthetic models predict. Thus, we conclude that the bumpy LCs oberved in luminous and superluminous SNe may not necessarily always originate from the variable thermalization of the magnetar spin-down energy. Several different mechanisms such as a CSM interaction could be active in making bumpy LCs in luminous and superluminous SNe.

\section*{Acknowledgements}
The authors thank the Yukawa Institute for Theoretical Physics (YITP) at Kyoto University, where this work was initiated during the YITP workshop YITP-T-21-05 on "Extreme Outflows in Astrophysical Transients."
We thank Matt Nicholl for providing the data of SN~2015bn and Sebastian Gomez for providing the data of SN~2019stc.
T.J.M. is supported by the Grants-in-Aid for Scientific Research of the Japan Society for the Promotion of Science (JP18K13585, JP20H00174, JP21K13966, JP21H04997).
This work of K.M. is supported by the NSF Grant No.~AST-1908689, No.~AST-2108466 and No.~AST-2108467, and KAKENHI No.~20H01901 and No.~20H05852.
The work by S.I.B. on SNIc studies is supported by RFBR-DFG project 21-52-12032 and on the code STELLA by grant RSF 19-12-00229.
Numerical computations were in part carried out on PC cluster at Center for Computational Astrophysics (CfCA), National Astronomical Observatory of Japan.

\section*{Data Availability}
The data underlying this article will be shared on reasonable request to the corresponding author.
 


\bibliographystyle{mnras}
\bibliography{mnras} 







\bsp	
\label{lastpage}
\end{document}